\documentclass[prfluids,12pt,tightenlines,showpacs,nofootinbib,floatfixlongbibliography]{revtex4-2}
\usepackage{epsfig}
\usepackage{amsmath,amssymb}
\usepackage{graphicx}
\usepackage{epsfig,latexsym,array,booktabs}
\usepackage{colordvi}
\usepackage{xcolor,xr}
\usepackage{orcidlink}

\newcommand{\be}{\begin{equation}}
\newcommand{\ee}{\end{equation}}
\newcommand{\ba}{\begin{eqnarray}}
\newcommand{\ea}{\end{eqnarray}}
\newcommand{\benn}{\begin{displaymath}}
\newcommand{\eenn}{\end{displaymath}}
\renewcommand{\d}[1]{{\rm d}#1}

\renewcommand{\vec}[1]{\mbox{\boldmath $#1$}}

\newcommand{\svec}[1]{\mbox{\scriptsize \boldmath ${#1}$}}
\newcommand{\mat}[1]{\mbox{\boldmath ${\rm #1}$}}
\newcommand{\e}[1]{\mbox{e}^{#1}}
\newcommand{\ii}{\imath\,}  
\newcommand{\wt}[1]{\widetilde{#1}}

\newcommand{\inttwo}{\int \!\!\! \int}

\reversemarginpar
\begin{document} 
\title{Faraday waves covered by a viscoelastic sheet}
\author{Hanna Pot\,\orcidlink{0009-0006-6275-7721}}
\affiliation{
Department of Maritime and Transport Technology, Delft University of
Technology, Delft, The Netherlands}
\author{Bram Christiaens}
\affiliation{
Laboratory for Aero- and Hydrodynamics, Delft University of Technology
and J.M. Burgers Centre for Fluid Dynamics, Delft, The
Netherlands}
\author{Willem van de Water\,\orcidlink{0000-0002-1054-9780}}
\email[Corresponding author: ]{w.vandewater@tudelft.nl}
\affiliation{
Laboratory for Aero- and Hydrodynamics, Delft University of Technology
and J.M. Burgers Centre for Fluid Dynamics, Delft, The
Netherlands}
\date{7 May 2026}
\begin{abstract}
The hydroelastic response of free floating viscoelastic covers is
measured using Faraday waves on the surface of a vertically
oscillated fluid layer. 
We systematically vary the thickness $d$ of the covers to investigate
its effect on the hydroelastic dispersion relation, the damping and
the isotropy of the waves. Compared to bare fluids, the wave patterns
are disordered. Various methods are explored to define and analyze
the wavelengths, the isotropy, and shape of the waves.
We find a significant difference between the measurements and the
theoretical dispersion relation.  Over all thicknesses $d$, this is
explained by an increase in the in-plane membrane tension, which
scales with $d^{3/2}$. 
Covering waves also has a large efect on their damping. Only for 
thin covers ($d = 20\: \mu{\rm m}$) the onset amplitude (and thus the
damping) can be explained by dissipation in the bulk and in the
boundary layer of the water beneath the cover.  
The same was found for bare water due to the presence of an immobile
surface layer.
Lastly, we find a large effect of the membrane on the ampitude of the
waves, which we attribute to nonlinear wave interaction.
\end{abstract}
\maketitle
%
%...................................................................
%...................................................................
%...................................................................
\newpage
\section{Introduction}
Covering the surface of a fluid with a viscoelastic membrane can
change the properties of free surface waves dramatically: it
increases the damping of waves, and changes their dispersion
relation.  The motion of the membrane is resisted by tension and
bending rigidity.
One-dimensional progressive waves, with elevation $\zeta(x, t) = a
\cos(k \: x - \omega t)$, then obey the dispersion relation 
\be
    \omega^2 = \frac{\rho \: g \: k + T \: k^3 +
    D \: k^5}{\rho \coth (k H) + \rho_s \: d\: k}, 
\label{eq.disprel.1}    
\ee
with $k$ the wavenumber, $H$ the depth of the fluid, $\omega$ the
frequency, and $D$ the bending rigidity of the membrane.  The tension
$T$ is composed of $\sigma_s$ due to the membrane, while the water
surface tension contributes $\sigma$: $T = \sigma + \sigma_s$.  The
bending rigidity follows from the membrane elasticity $E$ as 
\be
   D =\frac{E \: d^3}{12 (1 - \nu^2)},
\label{eq.rigid}   
\ee
with $\nu$ the Poisson ratio. In most cases the depth $H$ of the
fluid is much larger than $k^{-1}$ so that $\coth (k H) \approx 1$.

Research on these waves is motivated by the growing interest in
offshore engineering applications such as offshore floating solar
\citep{Trapani2013, Liu2024, Zhang2022}, and wave-sea ice
interactions in a geophysical context. In these fields, the
mechanical behavior of the cover is relevant since it provides
estimates of structural failure or ice breakup \citep{Bennetts2022}.
Insight in hydro-structural damping is required for wave penetration
forecasting \citep{Bennetts2022, Sree2018}, and eventual
environmental effects due to changes in vertical ocean mixing and
transport \citep{Benjamins2024, Denis2025}. 

The floating structures can be characterized by dimensionless
parameters such as the structure length $L$ over the wavelength
$\lambda$, and the ratio $L / \lambda_c$, with $\lambda_c$ a
characteristic length: a measure of hydroelastic stiffness
\cite{Zhang2022, Suzuki2006}. 
In engineering structures or sea-ice covers, flexure stiffness
prevails, and $\lambda_c$ is set by the balance between gravity and
flexure, $\lambda_c = 2\pi [{D}/({\rho g})]^{1/4}$.  If, on the
other hand, tension prevails, $\lambda_c = 2\pi [{T}/{(\rho
g})]^{1/2}$ \citep{Domino2018}.  Both definitions are suggested by
Eq.\ (\ref{eq.disprel.1}).  For the $200\: \mu{\rm m}$ covers in our
experiments and tension set by the surface tension of water, $T =
\sigma$, $\lambda_c \approx 10^{-2}\: {\rm m}$ for both tension and
flexure, so that the parameters of our experiments
are those of ``very large floating structures''.

Eq. \ref{eq.disprel.1} is derived from the Kirchhoff thin plate
equation \cite{Ventsel2001}.  This theory assumes that the plate
thickness $d$ is much smaller than the lateral dimension, that the
deflections are small, $\zeta \ll d$, and that shear deformation
through the thickness is absent. An additional in-plane tension 
can be included for pre-tension or compression. 
Fluid is coupled to the plate via the (Bernouilli) pressure, while
the linearized kinematic condition at the bottom surface of the
membrane reads $u_z = \partial \zeta / \partial t$.  The pressure
jump between the fluid and the sheet is caused by the tension which
scales with the curvature $\partial^2 \zeta/\partial x^2$ and 
bending, which scales with $\partial^4 \zeta/\partial x^4$ and the
bending rigidity. Because $d \ll \lambda$, the mass of the the cover
is negligible in most cases. In the absence of a viscoelastic cover
($D = 0$ and $\sigma_s = 0$), the dispersion relation Eq.\
(\ref{eq.disprel.1}) follows from the potential flow in the bulk of a
fluid and the linearized kinematic condition of the surface.

Thus, the dispersion relation Eq.~(\ref{eq.disprel.1}) involves
potential flow of the fluid, combined with an equation for the
pressure jump caused by flexure of the cover: the Kirchhoff thin plate 
equation \cite{Ventsel2001}. What is missing in this approach is the
dissipation inside the surface cover, vortical fluid motion near the
surface and the direct interaction between flow and viscoelastic
membrane through the effects of shear in the boundary layer on the
surface.

Studying the dynamic interactions of large floating structures is
often done on laboratory scale, in which conditions and dispersion
relations (or membrane properties) can easily be varied. This has
resulted in many recent studies of covered capillary-gravity waves.

%.................................................. terug naar Willem
The spectrum of turbulent surface waves under sheets with different
tension was studied by \citet{Vernet2025} who concluded the
applicability of a thermodynamic description. With wave amplitude $a
\approx 1\:{\rm mm}$, and wave steepness $k \: a \approx 0.02$,
bending modes of the used sheet (thickness $d = 500\:{\rm mm}$) were
neglected. 
This work was a follow up of \citet{Deike2013} who emphasized the
relevance of mode conversion. The suggestion is that for steep waves,
bending modes of the membrane(terms proportional to $k^5$ in Eq.\
\ref{eq.disprel.1}) may induce stretching through nonlinearity,
adding to terms proportional to $k^3$, and thus may augment the
surface tension $\sigma \rightarrow \sigma + \sigma_a$, 
\be
    \sigma_a = c \: E \: d \: (a \: k)^2,
\label{eq.nonlin}    
\ee
with $a \: k$ the wave steepness ($a \: k \approx 0.06)$, and with
the constant $c$ adjusted to $c \approx 3$.  The $d = 350 \: \mu{\rm
m}$ elastic membrane was clamped at its circumference, resulting in
circular (Bessel) modes whose wavelength depended on wave amplitude
\citep{Deike2013}.  A turbulent state resulted from excitation with
random noise in time, applied to a point on the membrane.  The mode
conversion effect, Eq.\ (\ref{eq.nonlin}), was inspired by elegant
work on an elastic string by \citet{Legge1984}.  How transverse waves
lead to longitudinal waves through nonlinearity has been discussed by
\citet{Peake2006}.

Nonlinear three-wave interaction of hydroelastic waves, a key
ingredient of a turbulent cascade \citep{Zakharov93}, was
demonstrated by \citet{Deike2017}.  However, three-wave interaction
is also the case for capillary waves on a bare, uncovered fluid
\citep{Zakharov67,Nazarenko2011} where it leads to wave turbulence at
strong enough driving.  

One-dimensional hydroelastic waves, stirred in a point on a
free-floating membrane were discussed by \citet{Biot2019}.  A large
variation of membrane thicknesses, $d = 50 \ldots 258 \: \mu{\rm m}$,
caused the bending modulus $D$ to vary over two orders of magnitude,
so that all dynamic effects embodied in Eq.\ \ref{eq.disprel.1},
namely gravity, surface tension and bending came into play.
With a wave amplitude $a \approx 0.2 \:{\rm mm}$, we conclude a large
wave steepness $a \: k \approx 0.2$, but no mention was made of mode
conversion.

Covering waves with a viscoelastic sheet changes the dispersion
relation, and thus the index of refraction of waves.  This was
studied by \citet{Domino2018} on $20 \ldots 800 \: \mu{\rm m}$ thick
sheets, excited in a point.  With a typical wave amplitude $a \approx
1 \: \mu{\rm m}$, a very small wave steepness ($ a\:k \approx 10^{-3}
\ldots 10^{-4}$) results.

For laboratory-scale experiments with gravity-dominated waves the
model scales often remain large, ${\cal O}(1 - 10 \:{\rm m})$,
requiring substantial wave tank sizes and cost, while only a few
models and a narrow parameter space can be tested \citep{Michele2023,
Kristiansen2022, Pot2025, Meylan2015, Montiel2013}.  
In this paper we describe a laboratory experiment involving Faraday
waves (\citet{Faraday1831}) covered by a visco-elastic membrane that
explores (dimensionless) parameter spaces relevant for large--scale
applications.  

Faraday waves arise on the surface of a fluid layer that is
vertically oscillated with frequency $F$.  Excitation is not through a
localized wavemaker, but through the modulation of gravity, which
distinguishes this experiment from others 
\citep{Deike2013,Deike2017,Domino2018,Biot2019}.
The Mathieu equation is the mathematical context of Faraday waves.
Solutions are harmonics (multiples of $F$) and subharmonics (odd
multiples of $F/2$).  They are amenable to Floquet analysis.  The
first resonance is the subharmonic at $F/2$.  
When the acceleration amplitude $a_s$ is increased from zero, Faraday
waves first appear when $a_s$ surpasses a threshold value $a_{cr}$
which is determined by viscous damping. The distance to threshold is
expressed by the parameter $\epsilon$, 
\be
    \epsilon = \frac{a_s}{a_{cr}} - 1, 
\label{eq.reduced}
\ee
so that $\epsilon = 0$ exactly at onset.  Due to its subharmonic
nature, the Faraday wave amplitude grows proportional to
$\epsilon^{1/2}$ above threshold.

In Sec.\ \ref{sec.methods} we describe the experimental setup, with
typical wave fields illustrated in Fig.\ \ref{fig.surf}.  The
measurement of wavenumber spectra, wave isotropy and wave crestedness
is discussed in Sec.\ \ref{sec.charac}.  Section\ \ref{sec.results}
presents results of dispersion relations and wave damping.  Nonlinear
mode conversion, embodied by Eq.\ (\ref{eq.nonlin}), is discussed in
Sec.\ \ref{sec.nonlin}.  The context of three-wave interaction, with
a few illustrative results, is sketched in Sec.\ \ref{sec.three}.

%...................................................................
%...................................................................
\section{Methods}
\label{sec.methods}
The cylindrical fluid container has a diameter of 150~mm and is
driven by a shaker (Br\"{u}el \& Kjaer 4808) with amplifier
(Br\"{u}el \& Kjaer 2719). The oscillations are controlled by a
multifunction synthesizer (NF Electronic Instruments 1930A) and
measured using an accelerometer connected to a signal conditioner
(PCB Model 482B11 ICP), which provides the acceleration amplitude up
to $\pm 1\%$.
The fluid depth is 10 $\pm$ 0.8 mm. The ambient temperature of the
laboratory and the fluid is 20 $\pm$ 2 $^\circ$C.

The surface gradient field is measured using the synthetic schlieren
technique \citep{Moisy2009}.  A random dot pattern (dot diameter
$0.25\: {\rm mm}$) is printed on transparent paper and affixed to the
underside of the container as an optical reference.  To enhance the
speckle contrast it is illuminated from below by an circular array of
light emitting diodes ($70 \: {\rm mm}$ diameter).
A camera (Basler a2A1920 2.3MP with Nikkor 24-85 mm lens) is mounted
800 mm above the container, allowing an approximately paraxial
reconstruction of the surface gradient field.  Image acquisition is
synchronized with the excitation signal via the sync-out of the
frequency synthesizer, and routed through a pulse rate divider which
ensures phase-locked recording.  With the divider set at a fixed odd
number, both half cycles of the Faraday waves are captured using a
camera exposure time of $2 \: {\rm ms}$.

Circular samples of transparent elastic films (Elastosil Film 2030,
Wacker Chemie AG) with diameters of 120~mm are free floating on the
fluid, leaving an average space of 15~mm with the walls. The samples
have thicknesses of $20, 50, 100$ and $(200 \pm 2\%) \: \mu {\rm m}
$, and a material density of $1.075 \pm 0.025 \: {\rm g \: cm}^{-3}$,
both by the manufacturer's statement.
The manufacturer's stress-strain test for small strains reports a
Young's modulus of 0.58 $\pm$ 0.02 MPa. With a static hydroelastic
compression test \citep{Vella2004} this is verified as 0.6 $\pm$ 0.2
MPa. A Poisson ratio of 1/2 is assumed.

To validate the experimental setup and the analysis procedures, a
fluid-only experiment with silicone oil is performed.  The results
are shown in the appendix.
Because the elastic films do not float on the silicone oil and are
chemically unstable when in contact with the oil, demineralised water
(Sigma-Aldrich, $\rho = 1000 \: {\rm kg}\:{\rm m}^{-3}$, $\eta = 1.00
\: {\rm m Pa s}$) is used for the hydroelastic experiments. 
The water is refreshed at the start of each day to minimize
contamination. During the experiments, the sheets remained in a
stable position, without making contact with the side walls of the
tank. Trapped air bubbles and water drops are removed prior to
testing.  Six wave amplitudes are tested at each frequency. The
steepness of the hydroelastic tests remained smaller than $a \: k =
0.2$ for all waves, and the steepness of the fluid-only tests smaller
than $a \: k = 0.4$. The Kirchhoff thin plate assumptions are
considered valid for $a \: k \leq 0.1$, while $0.1 < a\: k < 0.2$ is
known as a transition regime \citep{Ventsel2001}. For fluids, $a \: k
> 0.2$ is considered as a strongly nonlinear wave. 
%

%--------------------------------------------------------------------
\begin{figure}[t]
\centering
\includegraphics[scale = 0.8]{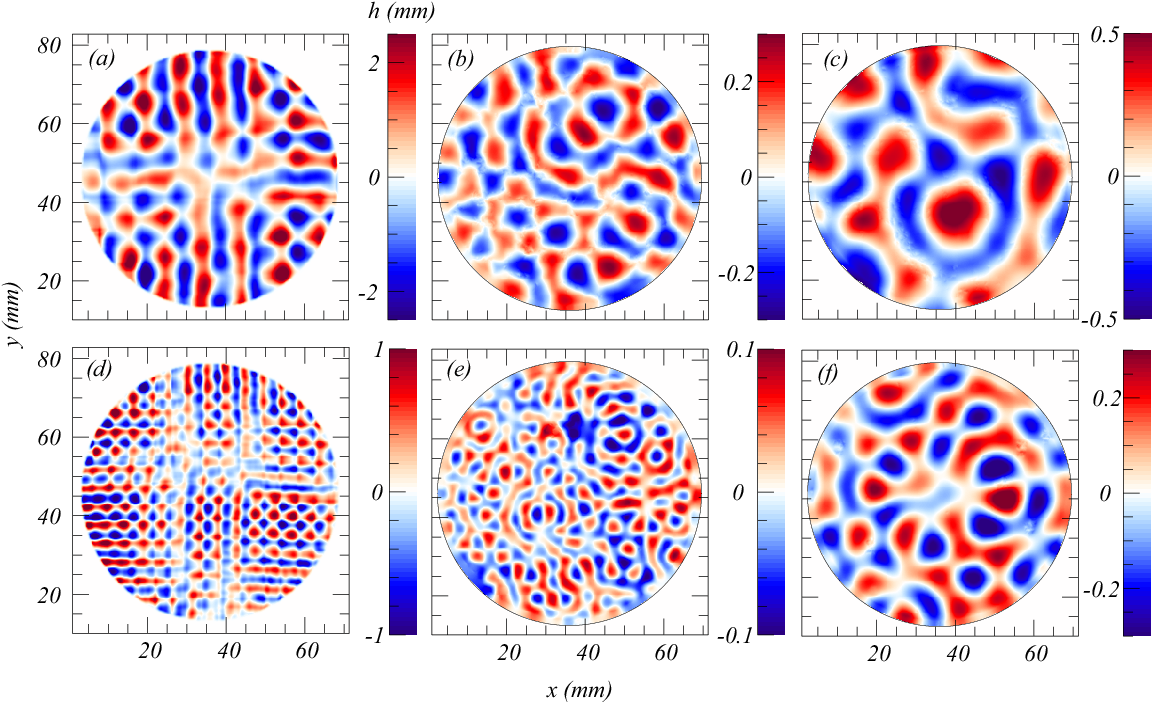}
\caption{
Snapshots of the elevation field $h(x, y)$.  Top to bottom: excitation
frequencies $F = 40, 90\:{\rm Hz}$, and left to right: uncovered water
($d = 0$), and two sheet thicknesses $d = 20, 200\:\mu{\rm m}$.
(a): $F = 40\:{\rm Hz}, d = 0,                \epsilon = 0.62$,
(b): $F = 40\:{\rm Hz}, d = 20 \: \mu{\rm m}, \epsilon = 0.53$,
(c): $F = 40\:{\rm Hz}, d = 200\: \mu{\rm m}, \epsilon = 0.32$,
(d): $F = 90\:{\rm Hz}, d = 0,                \epsilon = 0.32$, 
(e): $F = 90\:{\rm Hz}, d = 20 \: \mu{\rm m}, \epsilon = 0.79$,
(f): $F = 90\:{\rm Hz}, d = 200\: \mu{\rm m}, \epsilon = 0.18$,
where $\epsilon$ is the dimensionless driving amplitude [Eq.
(\ref{eq.reduced})]. }
\label{fig.surf}
\end{figure}
% see my notes
%--------------------------------------------------------------------

Each measurement run is preceded by registering an undisturbed
reference image of the background dot pattern.  Correlation with the
reference image gives the displacement field. The surface gradient
field then follows from simple geometric optics and the camera
calibration ($0.06 \: {\rm mm}$ per pixel).
Image correlation is performed with the open-source packages 
% \texttt{Ncorr} (Matlab, version~1.14.0.0) \citep{Blaber2015}
{\it Ncorr} \citep{Blaber2015}  and
% \texttt{OpenPIV} (Python) \citep{OpenPIV2021}.
{\it OpenPIV} \citep{Liberzon2021}.  For both packages an
interrogation window size of 24 pixels with a 20-pixel overlap is
used, ensuring continuous displacement analysis. 
The package {\it Ncorr} is applied to free-surface cases (both
silicone oil and deionized water), where large dot deformations and
occasional dot stretching occur, proving to be more robust under
these conditions, while {\it OpenPIV} offers faster processing and is
therefore used for the membrane cases.

The synchronously measured wave field varies randomly on a slow time
scale with a correlation time of approximately 70 periods of the
driving for sheets with $d = 200 \: \mu{\rm m}$.  The averages of all
experiments are done over 180 snapshots, which span $\approx 20$ and
$50$ correlation times at the lowest and highest excitation
frequency, respectively.  

A series of surface elevation snapshots is shown in Fig.\
\ref{fig.surf}, which illustrates the influence of the excitation
frequency and the dramatic influence of the viscoelastic cover.  
We nondimensionalize the excitation amplitude $a_s$ by its critical
value $a_{cr}$ as $\epsilon = a_s / a_{cr} - 1$.  At $\epsilon = 0$
the planform of wave patterns in uncovered fluids is determined by
the (circular) boundary conditions on the vertical container walls.
Above onset, $\epsilon \gtrsim 0.1$, the symmetry of the boundary no
longer matters, and the waves saturate into regular patterns: squares
in our case for uncovered water.  At selected frequencies
\citet{Westra2003} demonstrated quasicrystalline patterns in an
experiment.  Such a pattern has rotational symetry, but no
translational symmetry. At even larger driving amplitudes, $\epsilon
\gtrsim 1$, Faraday waves on bare fluids become disordered and
time-dependent.

The waves in the snapshots of Fig.\ \ref{fig.surf} are at relatively
weak driving.  The free surface waves on uncovered water exhibit
square symmetry, but covering them, even with the thinnest cover ($d
= 20\:\mu{\rm m}$) changes them dramatically. 

%...................................................................
%...................................................................
\section{Characterization of waves}
\label{sec.charac}
We characterize the waves in our experiment through the
instantaneous and time-averaged wavenumber spectra of the gradient
field, through the spatial symmetry of the gradient field in real
space, and, finally, through the ``crestedness'' of the elevation
field.

Our prime source of information is the surface height {\em gradient
field} of snapshots, $\nabla h = (\partial h/ \partial x,\partial h/
\partial y) \equiv (h_x, h_y)$, a result of the cross-correlation of
the imaged dot displacement field.  The elevation field $h(x, y)$
then follows from integrating the vector field $(h_x, h_y)$ over the
surface. Next, the mean elevation is set to 0, and slight
misalignments of the setup are corrected by fitting a plane to $h(x,
y)$ and subtracting it.  We quantify the surface amplitude as the
root mean square (rms) $\langle h^2 \rangle^{1/2}$, with an average
$\langle \cdots \rangle$ over the surface and over snapshots.  
%

%...................................................................
%...................................................................
\subsection{Spectra}
\label{sec.spectra}

For a better assessment of the surface wavelength, we combine the $x-$
and $y-$component of the gradient field into the complex field 
\be
   \wt{h}_{x,y} = h_x(x, y) + \ii h_y(x, y),
\label{eq.spec.1}   
\ee
with $\ii$ the imaginary unit.  Its Fourier transform is
\be
   {\cal F}(\wt{h}_{x,y}) =
   \inttwo_{-\infty}^\infty \e{\ii k_x x + \ii k_y y } \:
   h_x \: \d x \d y +
   \ii  
   \inttwo_{-\infty}^\infty \e{\ii k_x x + \ii k_y y } \:
   h_y \: \d x \d y.
\label{eq.spec.2}   
\ee
Practically we use zero padding to increase the wavenumber
resolution. The first term in Eq.\ (\ref{eq.spec.2}) can be partially
integrated over $x$, the second term can be partially integrated over
$y$ (assuming that $\nabla h$ is zero outside the domain of
interest), so that
\be
   {\cal F}(\wt{h}_{x,y}) =
   (-\ii k_x + k_y) 
   \inttwo _{-\infty}^\infty \e{\ii k_x x + \ii k_y y }\:
   h(x, y) \: \d x \d y.
\label{eq.spec.3}   
\ee
The energy spectrum is $S_{\svec{k}}(k_x, k_y) = k^2 | {\cal F}
(h(\vec{x})) |^2$.  
Because of the modulus, obviously $S_{\svec{k}}(-k_x, -k_y) =
S_{\svec{k}}(k_x, k_y)$, $S_{\svec{k}}(-k_x, k_y) = S_{\svec{k}}(k_x,
-k_y)$, so that a representation of $S_{\svec{k}}(k_x, k_y)$ in the
upper half wavenumber plane suffices.  The other half is related to
it via trivial symmetry.  
\footnote{This distinguishes our spectra from those in Fig.6 of
\citet{Deike2013}, which are space--time ($k - \omega$) spectra.
Since we sample synchronously with the excitation, our spectra are
effectively at $\omega = \pi \: F$, and multiples thereof.}.

For randomly oriented waves, the spectrum consists of rings in
wavenumber space. Multiplication with $k^2$, the consequence of
taking the gradient field, makes them stand out more clearly, which
facilitates measurement of the wavelength.

\begin{figure}[t]
\centering
\includegraphics[scale = 0.8]{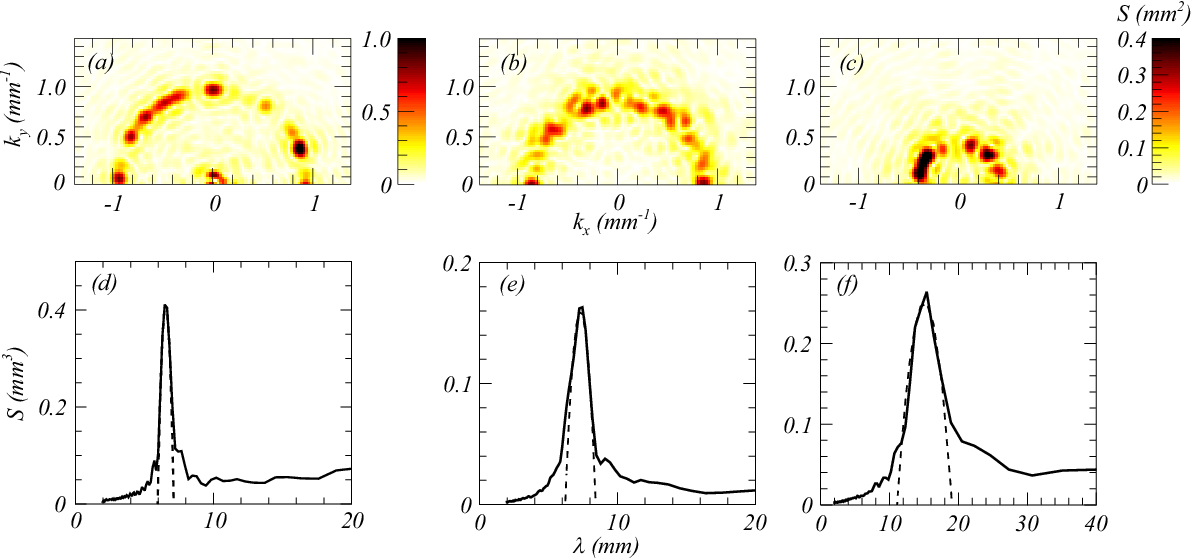}
\caption{
Panels (a-c): ensemble averaged energy spectra $S(k_x, k_y)$.  Panels
(d-f) Azimuthally integrated spectra $S$ as a function of $\lambda =
2 \pi / k$.  The dominant wavelength is the maximum of the spectra, its
uncertainty is (arbitrarily) set to the width of the peaks at 0.8
maximum height.
The dashed lines are quadratic fits for determining the center
wavelength and width.
Uncovered water in panels (a,d), membrane thickness $d = 20 \: \mu
{\rm m}$ in panels (b,e), and $d = 200 \: \mu {\rm m}$ in panels (c,
f). 
In all cases the driving frequency is $60 \: {\rm Hz}$.
}
\label{fig.iso1}
\end{figure}

Before discussing the isotropy of surface snapshots and their
spectra, we first determine their dominant wavenumber $k_p$.  To this
aim we average snapshot spectra over the azimuth angle $\phi$ and
over realizations $\left\langle \cdots \right\rangle$, 
\be
   S(k) = \left\langle \frac{1}{\pi} \int_0^\pi S(k, \phi) \: \d \phi
\right\rangle.
\ee
The results $S(k)$, shown in Fig.\ \ref{fig.iso1}, display sharp
maxima, which define $k_p$.  The shape of the spectra $S(k)$ is
determined by the convolution of the Fourier transform of the
measurement domain, and that of the waves.  We expect that the latter
is influenced by the viscosity of the fluid and viscoelastic
properties of the membrane.  However, it is obscured by the finite
size $L$ of the domain, leading to a wavelength spread $\Delta
\lambda \approx \lambda^2 / L$.

%--------------------------------------------------------------------
%--------------------------------------------------------------------

%--------------------------------------------------------------------
\subsection{Surface isotropy}
\label{sec.isotropy}

Since thicker membranes may be prone to bending resistance and
consequent anistropy, we explore various methods to quantify the
isotropy of the waves.  We first establish that the instantaneous
spectra (wavenumber spectra of snapshots) are concentrated on
circular arcs. To this aim, these spectra were fitted to ellipses,
\be
   S_{\svec{k}}(k_x,k_y) = S_e^{a,b}(\phi), 
\label{eq.fellips}   
\ee
with $k_x = a\: \sin \phi$ and $k_y = b (1 - k_x^2/a^2)^{1/2}$. The
measured ratio of principal axes lengths $a / b$ fluctuates, but over
time, their average is equal to one: there is not a significant
difference with a circular arc $a = b = k$.  We will, therefore,
denote these spectra as $S(k, \phi)$.  
Clearly, the parametric excitation results in waves that are
homogeneous and do not have a preferred spatial orientation, unlike
the waves in experiments that use localized wavemakers
\citep{Deike2013,Deike2017,Domino2018,Biot2019}.

%--------------------------------------------------------------------
\begin{figure}[t]
\centering
\includegraphics[scale = 0.8]{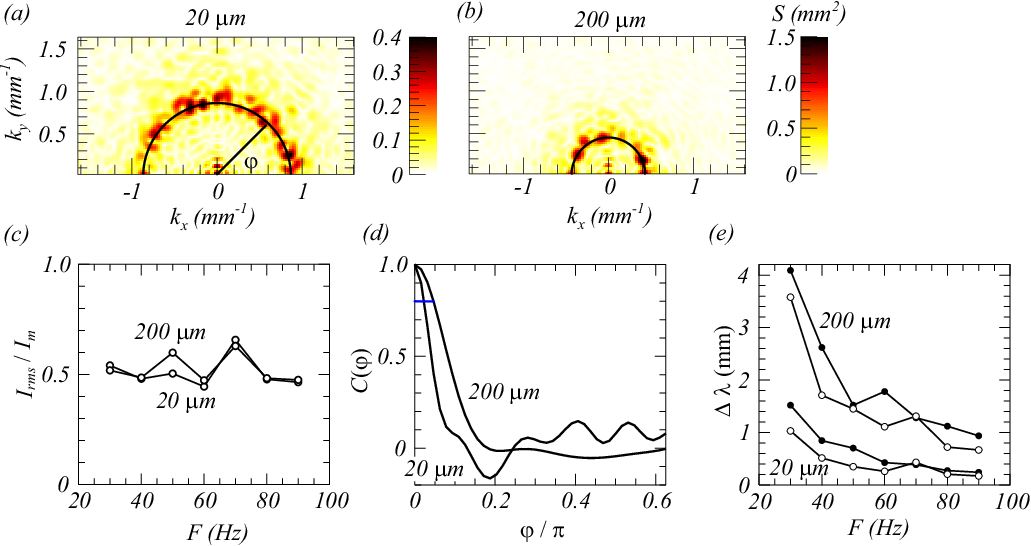}
\caption{
(a,b) Spectra of snapshots of the elevation field $h(x, y)$.  The
driving frequency is $F = 60 \: {\rm Hz}$, the sheet thickness is $d
= 20 \: \mu{\rm m}$ and $d = 200 \: \mu{\rm m}$, for panels (a) and
(b), respectively.  The lines are fits of ellipses, $k_x^2/a^2 +
k_y^2 / b^2 = k^2$. We trace the spectral energy $I(\phi) = S_e(k,
\phi)$ along these ellipses.  
(c) Normalized variation $I_{\rm rms} / I_{\rm m}$ of $I(\phi)$ for
two different membrane thicknesses.  
(d) Correlation $C(\phi)$ of $I(\phi)$ [Eq.\ (\ref{eq.corrphi})]. We
define the width of the correlation as the angle $\phi$ where $C =
0.8$ (indicated by a blue line).  
(e) Comparing the width of the azimuthal correlation (open cicles) to
the radial width (dots) of the azimuthally integrated spectra (see
Fig.\ \ref{fig.iso1}).
}
\label{fig.iso2}
\end{figure}
%--------------------------------------------------------------------

The distribution of spectral energy $I(\phi) = S(k_p, \phi)$ along
the arcs [Eq.\ (\ref{eq.fellips})], $(\phi \in [0, \pi], k = k_p$), 
shown in Fig.\ \ref{fig.iso2}, has remarkable statistical properties.
Snapshot spectra are shown in Fig. \ref{fig.iso2}(a,b).  They are
characterized by lumped concentrations of wavenumbers which do not
point in a particular direction.  This contrasts results by
\citet{Deike2013} who find anisotropy caused by the point-excitation
in their experiments.  
The ratio of root-mean-square, $I_{\rm rms} = (\langle I^2 \rangle -
\langle I \rangle^2)^{1/2}$, and mean values, $I_m = \langle I
\rangle$, of $I(\phi)$ is shown in Fig.\ \ref {fig.iso2}(c), where
averages are done over $\phi$ and over snapshots.  The {\em relative}
$I_{\rm rms} / I_m$ appears independent of driving frequency and
membrane thickness, and is close to $I_{\rm rms} / I_m = 1/2$.
The fluctuations are further quantified by the normalized
self-correlation function
\be
    C(\phi) = \left[
    \left\langle I(\phi' + \phi) \: I(\phi') \right\rangle 
    - I_m^2 \right] / I^2_{\rm rms}.
\label{eq.corrphi}
\ee
Two correlation functions for $d = 20 \: \mu{\rm m}$ and $d = 200 \:
\mu{\rm m}$ are shown in Fig.\ref{fig.iso2}(d).  
While the radial width $\Delta k$ of the spectra $S(k, \phi)$ is
discussed in Sec.\ \ref{sec.spectra}, the angular width $k\: \Delta
\phi$ is the width of the correlation function $C(\phi)$.  As Fig.\
\ref{fig.iso2}(d) demonstrates, these widths are approximately the
same.  The circular blobs in the wavenumber plane suggest
concentrated wave packets with a well-defined wavelength.  In case of
wave patterns with square crystalline order [such as in Fig.\
\ref{fig.surf}(a,d)], the correlation function $C(\phi)$ displays
peaks at $\phi = \pi/2$.  In case of an $n-$fold symmetry, these
peaks would be located at $\phi = 2\pi / n$ \citep{Westra2003}.

%--------------------------------------------------------------------
\begin{figure}[t]
\centering
\includegraphics[scale = 0.8]{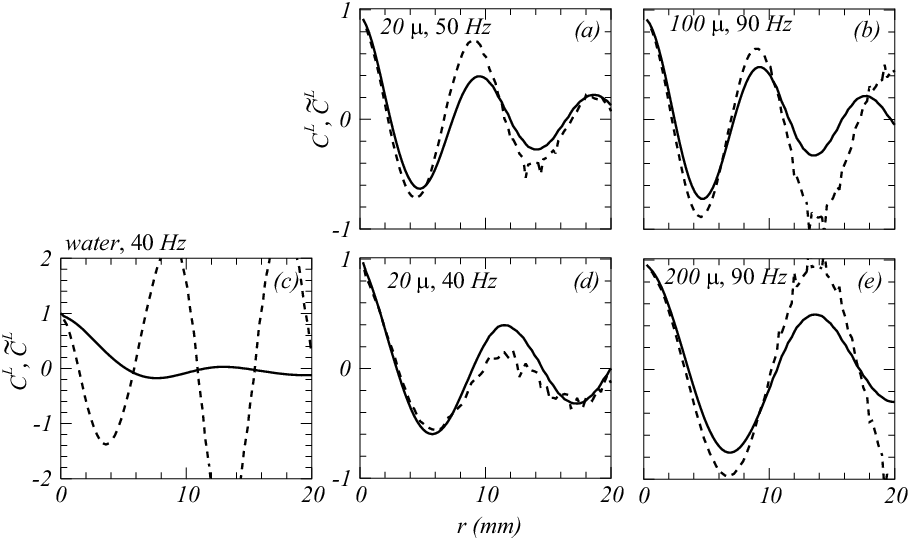}
\caption{
Isotropy relation for longitudinal and transverse correlation
functions.  
In the top row (panels a, b) we compare $d = 20\:\mu{\rm m}$ and $d =
100\:\mu{\rm m}$, with wavelengths $\lambda = 8.5\:{\rm mm}$ and
$8.4\:{\rm mm}$, respectively. In the second row (c,d,e) we compare
bare water, $d = 20\:\mu{\rm m}$ and $d = 200\:\mu{\rm m}$, with
wavelengths $\lambda = 7.9, 10.8$ and $11.8\:{\rm mm}$, respectively.
Full lines are the measured longitudinal correlations $C^L$,
dashed lines are the correlations $\wt{C}^L$ computed from the
measured $C^T$ and isotropy relation Eq.\ (\ref{eq.iso}).  
}
\label{fig.clt}
\end{figure}
%--------------------------------------------------------------------

While the spectral energy $S_{\svec{k}}(k_x, k_y)$ is still tied to
the coordinate axes, a more general measure of anisotropy starts from
the surface gradient field $h_{x, y}$.  There are two distinct ways
in which this fluctuating vector field can be analyzed: {\em
longitudinal}, when correlations of $h_x$ are measured in the
$x-$direction, and {\em transverse} when they are analyzed in the
$y-$direction, and vice-versa.  These arrangements are reflected in
the one-dimensional (1D) correlation functions
\be
   C_{\alpha \alpha}( r \: \vec{e}_\beta) = 
   \frac{
   \left \langle h_\alpha(\vec{x} + r \vec{e}_\beta)\: h_\alpha(\vec{x})
   \right \rangle -
   \langle h_\alpha \rangle^2}
   {
   \langle h_\alpha^2 \rangle - \langle h_\alpha \rangle^2
   },
\label{eq.corrlt} 
\ee
with $\alpha, \beta = x, y$ and averages $\langle \ldots \rangle$
over the measurement domain and over snapshots.  We call $C$ the {\em
longitudinal} correlation $C^L$ when $\alpha = \beta$, so $C^L(r) =
C_{x x}(r \vec{e}_x)$ and $C^L(r) = C_{y y}(r \vec{e}_y)$. This
correlation is equivalent to the usual 1D spectrum.  Conversely, the
{\em transverse} correlation $C^T$ has $\alpha \ne \beta$.  The
fields $h_x, h_y$ are gradient fields, and in case of isotropic waves
the correlation functions must satisfy \cite{Savelsberg2009}
\be
   C^L(r) = C^T(r) + r \: \frac{\d}{\d r} \: C^T(r).
\label{eq.iso}   
\ee   
Equation\ (\ref{eq.iso}) involves the measured quantities $C^L$ and
$C^T$, and its satisfaction reflects the degree of isotropy of the
surface waves.  From the measured $C^T$ and Eq.\ (\ref{eq.iso}) we
compute $\wt{C}^L$ and compare it to the actually measured $C^L$. 
In Fig.\ \ref{fig.clt} we compare $C^L$ and $\wt{C}^L$ for $d = 20
\mu$, $F = 40 {\rm Hz}$ and $d = 200 \mu$, $F = 90 {\rm Hz}$.  
As the details of the correlation function depend on the shape of the
measurement domain and the number of waves in the measurement domain,
we compare cases with the same wavelength but different thickness $d$
of the cover.  In this way we isolate the influence of $d$.  In Fig.\
\ref{fig.clt} we observe a clear anisotropy for the thickest ($d =
100, 200 \: \mu{\rm m}$) membranes compared to that with $d = 20 \:
\mu{\rm m}$.  As expected from Fig.\ \ref{fig.surf}, this measure
also highlights the strong anisotropy of the square wave pattern on
bare water.

%....................................................................
\subsection{Wave crests}

All other experiments on waves in fluids covered by membranes
involved localized wavemakers
\citep{Deike2013,Deike2017,Domino2018,Biot2019}, which caused waves
whose crests run parallel, i.e. one-dimensional (1D) waves.  It is
expected that the dynamics of hydroelastic waves depends on their
dimensionality as there is a larger bending penalty in case the
surface sheet is bent in two directions simultaneously.

%--------------------------------------------------------------------
\begin{figure}[t]
\centering
\includegraphics[scale = 0.8]{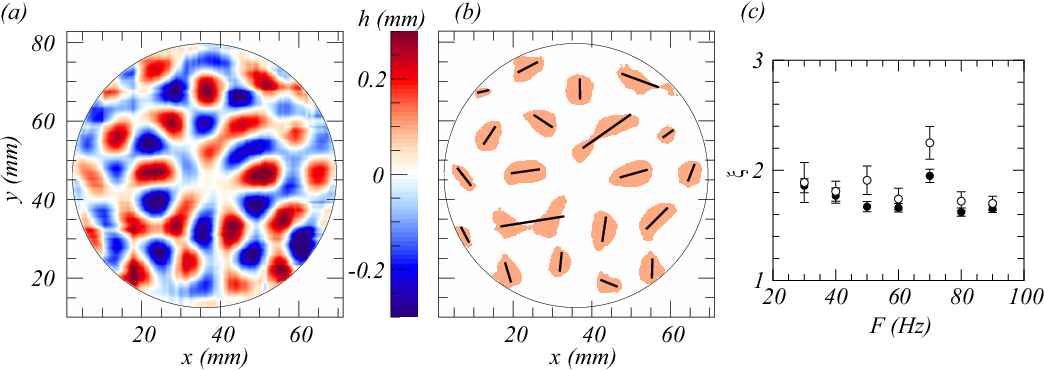}
\caption{
(a) Surface elevation $h(x, y)$ at $F = 90\:{\rm Hz}$ and sheet
thickness $d = 200 \: \mu{\rm m}$.
(b) Isolating wave crests at height $a_t = 0.1\:{\rm mm}$, the lines
indicate the longest principal axes.  
(c) Aspect ratio $\xi$ of crests at height larger than or equal to to
the rms amplitude, $a_t \ge a$, averaged over crests and snapshots.
Open circles: $d = 200 \: \mu{\rm m}$, dots: $d = 20 \: \mu{\rm m}$.
The error bars are the rms variation of $\xi$. 
}
\label{fig.blob}
\end{figure}
%--------------------------------------------------------------------

Wave crests can be recognized in Fig.\ \ref{fig.surf}, and the
question is if they are more 1D than 2D.  To provide a quantitative
answer we find all regions $h_i(\vec{x})$ (``blobs'') on the surface
with elevation $h_i \ge a_{t}$.  The aspect ratio $\xi$ of these
blobs is the ratio of their principal axes length.  The value of the
average $\xi$ reveals the dimensionality, it is one for 2D circular
blobs, while $\xi > 1$ for long-crested waves.

We find these axes by forming the matrix $\mat{M}$ with elements
\be
   M_{k l} = \inttwo (x_k - \bar{x}_k)\: (x_l - \bar{x}_l) \:
   h_i(x_k, x_l) \: \d x_k \: \d x_l,
\ee
where $x_k, x_l$ are the components of the vector $\vec{x}$, and
$\bar{\vec{x}}$ is the blob center of mass, 
\benn
   \bar{\vec{x}} = \inttwo \vec{x} h_i(\vec{x})\: \d \vec{x} \: \bigg/ 
    \inttwo h_i(\vec{x})\: \d \vec{x}.
\eenn     
The eigenvectors of $\mat{M}$ are the principal axes, the ratio of
its eigenvalues is the aspect ratio $\xi$ of the blob.  The results
for $d = 20 \: \mu{\rm m}$ and $d = 200 \: \mu{\rm m}$ are shown in
Fig.\ \ref{fig.blob}.  The threshold surface height $a_t$ was taken
equal to the rms wave amplitude $a$.  There is no significant
difference between the two sheets, neither is there a significant
influence of the chosen threshold surface height $a_t$.

Summarizing, covered Faraday waves are disordered spatially, but with
a well-defined wavelength.  Although they do not have a preferred
direction, they are slightly anisotropic.  Surprisingly, their crests
are elongated (aspect ratio $\xi \approx 2$), but the elongation is
independent of membrane thickness and excitation amplitude within our
experimental excitation amplitude range. 

%--------------------------------------------------------------------
\begin{figure}[t]
\centering
\includegraphics[scale = 0.8]{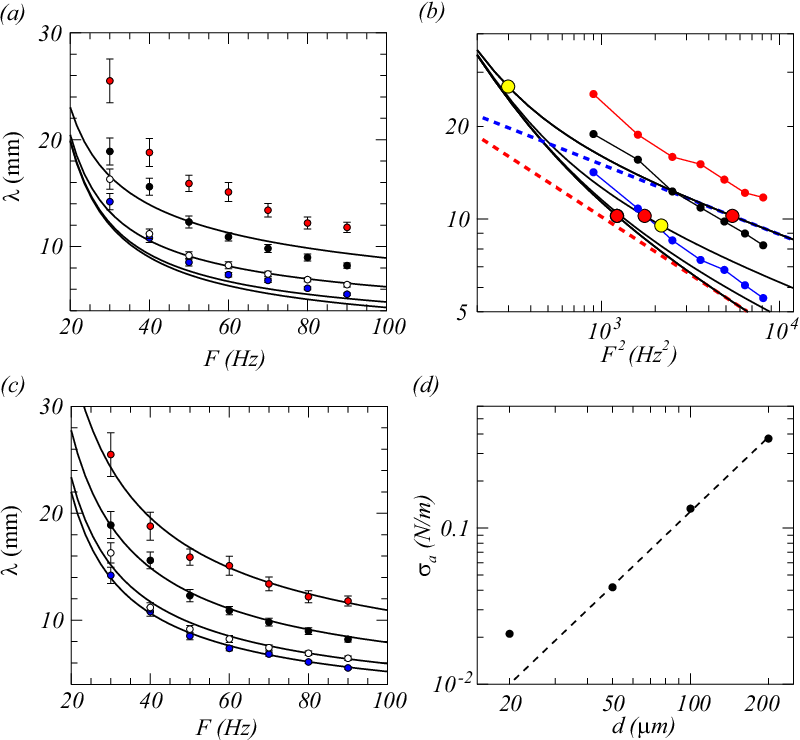}
\caption{
(a): Dots with error bars: measured dispersion relation of water
covered by a membrane with thickness $20 \: {\mu}{\rm m}$ (blue
dots), $50 \: {\mu}{\rm m}$ (white dots), $100 \: {\mu}{\rm m}$
(black dots) and $200 \: {\mu}{\rm m}$ (red dots).  Full lines:
dispersion relation Eq.\ \ref{eq.disprel.1}, using the fitted surface
tension for bare water, $\sigma_0 = 2.80\times 10^{-2} \:{\rm N}{\rm
m}^{-1}$ and measured membrane elasticity $E = 0.58\times 10^6 \:
{\rm N}{\rm m}^{-2}$.
(b): Same data as in panel (a), but now plotted in a way to elucidate
scaling behavior. The yellow dots indicate the cross-over from
stretching to bending modes.  The red dots indicate the cross-over
from gravity to stretching.
Dashed lines: fits to large $k$ (small $\lambda$) behavior, red
line(s): $\lambda \propto (F^2)^{-1/3}$, blue: $\lambda \propto
(F^2)^{-1/5}$.
(c): Same data as in panel (a).  Lines: Eq.\ \ref{eq.disprel.1}, but
with thickness dependent surface tension: $\sigma = \sigma_{0} +
\sigma_a$, with $\sigma_{0} = 2.8\times 10^{-2} \:{\rm N}{\rm
m}^{-1}$, the measured surface tension of water, and $\sigma_a$ as
shown in panel (d)
(d): Dependence of $\sigma_a$ on $d$ in a log-log plot; dashed line:
$\sigma_a \propto d^{1.6}$.
}
\label{fig.disp3}
\end{figure}
%--------------------------------------------------------------------

\section{Dispersion and damping of waves}
\label{sec.results}
\subsection{Dispersion}
\label{sec.dispersion}

The dispersion relation Eq.\ \ref{eq.disprel.1} relates the
wavenumber of waves to the wave frequency, $\omega = F / 2$.  In this
section it is compared to measured wavelengths using known values of
the physical parameters $\rho, \sigma$ and $D$, assuming that
$\sigma_s$ of the cover is absent.
The nominal surface tension of water is $7.275\times 10^{-2} \: {\rm
N}{\rm m}^{-1}$, however, in practice it is influenced by unavoidable
surface contamination.  Therefore, the value taken in our experiments
is $\sigma = \sigma_0 = 2.8\times 10^{-2} \: {\rm N}{\rm m}^{-1}$, as
is determined from a least squares fit of the dispersion relation.
A similar approach was followed in
\cite{Douady90,Domino2018,Biot2019}.
The bending rigidity $D$ of the membrane follows from the measured
elastic constant $E = 0.58\times 10^6 \: {\rm N}{\rm m}^{-2}$ [Eq.\
(\ref{eq.rigid})].
The comparison of the measured wavelengths to the prediction of the
dispersion relation is shown in Fig.\ \ref{fig.disp3}(a).  The
discrepancy is striking.   

The dispersion relation Eq.\ \ref{eq.disprel.1} predicts scaling of
the squared wave frequency $\omega^2 \propto k$ for gravity,
$\omega^2 \propto k^3$ for surface stretch, and $\omega^2 \propto
k^5$ for sheet bending.  In \ref{fig.disp3}(b) we show the measured
$\lambda$, versus $F^2$ in a log-log plot, which highlights these
scalings.
The cross-over from stretch to bending occurs at wavenumber $k =
(\sigma / D)^{1/2}$.  For larger larger frequencies (larger
wavenumbers) the measured wavelength would then scale as $\lambda
\propto (\omega^2)^{-1/5}$. As Fig.\ \ref{fig.disp3}(b) illustrates,
all measured wavelengths of the $d = 200 \: \mu{\rm m}$ cover should
show this scaling, and some experiments involving the $d = 100 \:
\mu{\rm m}$ cover. For the thinnest $d = 20 \: \mu{\rm m}$ sheet, all
scaling would be the stretch scaling, $\lambda \propto
(\omega^2)^{-1/3}$. Although the dynamical range of frequencies is
limited, this change of scaling can be observed in Fig.\
\ref{fig.disp3}(b).

To quantify the discrepancy in Fig.\ \ref{fig.disp3}(a), we set
$\sigma = \sigma_0 + \sigma_a$, and determine $\sigma_a$ in a least
squares procedure for each sheet thickness $d$.  As Fig.\
\ref{fig.disp3}(c) illustrates, this reproduces the measured
dispersion relations very well.  The result $\sigma_a$ of these fits
is shown in Fig.\ \ref{fig.disp3}(d); it appears that $\sigma_a
\propto d^{3/2}$, but we have no explanation for this heuristic fit.

%--------------------------------------------------------------------
\begin{figure}[t]
\centering
\includegraphics[scale = 0.8]{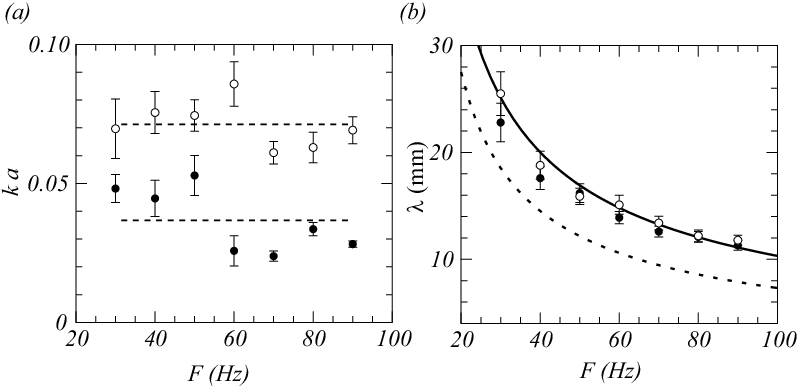}
\caption{
(a) Wave steepness $k\:a$ for two excitation acceleration amplitudes.
The dashed lines indicate the average $k \: a$; the related scaling
is used in the dispersion relation shown in panel (c).
(b)
Measured dispersion relations for the excitation amplitudes
$a_s$, of panel (a).
Full line: dispersion relation Eq.\ \ref{eq.disprel.1} using $\sigma
= \sigma_0 + \sigma_a$ with $\sigma_a = 0.38 \: {\rm N}{\rm
m}^{-1}$.  Dashed line: dispersion relation with {\em scaled}
$\sigma_a = 0.10 \: {\rm N}{\rm m}^{-2}$.
}
\label{fig.disp4}
\end{figure}
%--------------------------------------------------------------------

%--------------------------------------------------------- Nonlinearity
\subsection{Nonlinearity}
\label{sec.nonlin}

Deformations of the membrane may induce additional tensions due to
nonlinearity.  This may explain the $\sigma_a$ as we found in Fig.\
\ref{fig.disp3}(c). The effect is demonstrated by \citet{Deike2013}
who found a shift of the wavelength as a function of the forcing
amplitude for a sheet clamped at its circumference. Nonlinearity is
gauged by the wave steepness $k\: a$.  
In our experiment we can tune the wave amplitude $a$ through
variation of the excitation amplitude $a_s$.  At the highest
amplitude $a_s$, $k\:a \approx 0.07$; at the lowest amplitude $k\:a
\approx 0.04$.  Since the additional tension depends quadratically on
the steepness, this should result in a measurable change of the
wavelength, if Eq.\ (\ref{eq.nonlin}) holds.  The effect should be
largest for the thickest sheet due to its proportionality to $d$.
Since the constant $c$ is not known, we check Eq.\
(\ref{eq.nonlin}) in a {\em relative} sense: from a fit we find
$\sigma_a \approx 0.38$ at the highest excitation amplitude, which
then should reduce to $\sigma_a \approx 0.10$ at the lowest
amplitude.  Figure\ \ref{fig.disp4}(b) demonstrates that this is not
the case: the nonlinear effect embodied by Eq.\ (\ref{eq.nonlin})
is not measurable.  

Notice that the steepness in Fig.\ \ref{fig.disp4}(a) varies with
frequency, but it has large error bars due to the large variations of
the wave amplitude together with the wavelength uncertainty. In Fig.\
\ref{fig.disp4}(a) we quantify the steepness change by its frequency
average, which results in the dashed dispersion curve in Fig.\
\ref{fig.disp4}(b).  Alternatively, we could have computed $\sigma_a$
{\em per frequency}.  In any case our conclusion remains unchanged.

Summarizing, we find a strong dependence of the measured dispersion
relation on the membrane thickness.  It is consistent with a change
of the surface tension, but it can not be explained by the nonlinear
mode conversion Eq.\ (\ref{eq.nonlin}).

%--------------------------------------------------------------------
\begin{figure}[t]
\centering
\includegraphics[scale = 0.8]{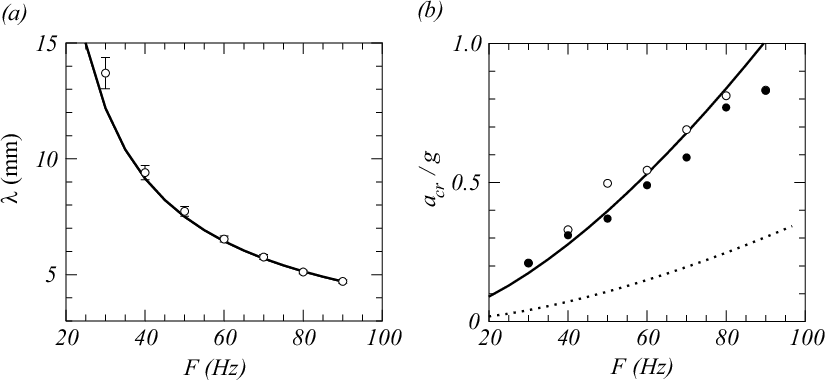}
\caption{
(a) Open circles: measured dispersion relation of uncovered water,
the error bars follow from the spectral widths, as illustrated in
Fig.\ \ref{fig.iso2}.  Two overlapping full lines : predicted using
full hydrodynamics \citep{Kumar94}, and the phenomenological model
with damping $\gamma_b + \gamma_s$, respectively.  These results are
indistinguishable. The used surface tension is $\sigma = \sigma_0 = 
2.8\times 10^{-2} \: {\rm N}{\rm m}^{-1}$.
(b) (Dimensionless) critical excitation acceleration amplitude
$a_{\rm cr} / g$. Open circles: uncovered water, dots: covered by a
membrane with thickness $d = 20 \: {\mu}{\rm m}$.  Full line using
the phenomenological model with both bulk- and surface damping,
$\gamma_b + \gamma_s$.  Dashed line: as computed using ``the full
hydrodynamic approach'' by \citet{Kumar94}, which includes vorticity
near the surface, but lacks the surface damping $\gamma_s$. 
}
\label{fig.disp1}
\end{figure}
%--------------------------------------------------------------------

%============================================================ damping
\subsection{Damping}
\label{sec.damping}
Apart from the wavenumber, also the critical excitation amplitude
$a_{cr}$ is accessible in an experiment. It is determined by wave
damping and can be compared to theoretical predictions.
A complete hydrodynamical description of Faraday waves on the
interface between two immiscible fluids is given by \citet{Kumar94}.
It involves detailed Navier Stokes equations for viscous fluid flow
and appropriate boundary conditions: there is a pressure jump at the
interface because of interfacial tension and the gradient of vertical
flow, while a no-slip boundary condition is assumed on the container
walls.  In our case we take the upper fluid as air, so that,
effectively, the surface is completely mobile and the water-air
boundary condition becomes that of no strain, $\partial u/ \partial z
|_{z = 0} = 0$.  
\citet{Kumar94} demonstrate that their results for $a_{cr}$ are very
close to an approach involving a Mathieu equation with a
phenomenological damping factor \citep{Batchelor00}
\be
    \gamma_b = -2 \: \nu \: k^2,
\label{eq.damp.bulk}
\ee
which follows from the dissipation computed using a potential flow
field in the bulk, and where $\nu$ is the kinematic viscosity.  This
phenomenological damping misses the dissipation due to the rotational
flow near the surface which is contained in the full hydrodynamic
approach \citep{Kumar94}.  However, the interaction between harmonics
of the Floquet method is retained.

The surface contamination has a strong effect on the damping of
waves.  With mobile surfactant molecules, damping results from
Marangoni effects.  With immobile molecules, there is a no-slip
condition at the surface, $u|_{z = 0} = 0$.  In the latter case, 
dissipation also occurs in the thin boundary layer below the surface.
The associated damping factor then follows from Stokes' second
problem \citep{Lighthill1978},
\be
    \gamma_s = -\frac{k}{2} \left(\nu \: \omega / 2 \right)^{1/2}.
\label{eq.damp.surf}    
\ee

The Mathieu equation without damping predicts the occurence of waves
at vanishing excitation amplitude $a_s$.  With damping, a finite
critical amplitude $a_{cr} > 0$ is required where subharmonic waves
first emerge.
\footnote{For very shallow fluid layers, {\em harmonic} waves may
emerge first \citep{Muller1997}.}
The amplitude of waves grows with time constant $\tau \propto
\epsilon^{-1}$ after the excitation is turned on.  It becomes
infinitely long at onset $a_s = a_{cr}$. Practically, the critical
amplitude $a_{cr}$ is determined as the amplitude where waves emerge
after a waiting time of minutes. Therefore, it is always
overestimated.  

%--------------------------------------------------------------------
\begin{figure}[t]
\centering
\includegraphics[scale = 0.8]{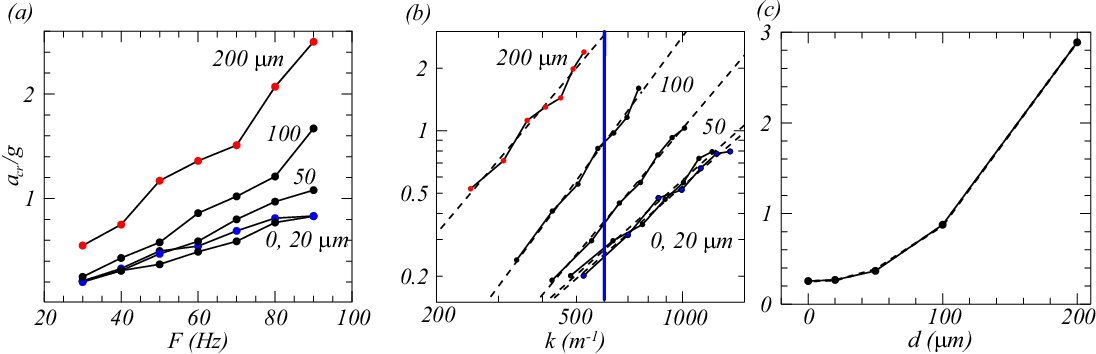}
\caption{
(a) Measured dimensionless critical acceleration amplitude $a_{cr} / g$
for uncovered water and viscoelastic covers with thickness $20$ (blue
dots), $50$, $100$ and $200 \: {\mu}{\rm m}$ (red dots).
(b) The data of panel (a), plotted as a function of wavenumber in a
log-log plot.  Dashed lines: $(a_{cr} / g) \propto k^b$, with $b = 1.5,
1.5, 2.0, 2.3$ and $2.0$ for bare water, and $d = 20, 50, 100, 200 \:
\mu{\rm m}$, respectively.
(c) Dimensionless critical excitation amplitude at $k = 600 \: {\rm
m}^{-1}$ [the blue line in panel (b)] as a function of membrane
thickness.  The dashed line is a fit $a_{cr} \propto d^2$. 
}
\label{fig.disp5}
\end{figure}
%--------------------------------------------------------------------

Damping affects both the dispersion relation as well as the critical
onset amplitude.  First, Fig.\ \ref{fig.disp1}(a) demonstrates that
the theoretical dispersion relation, computed from the Mathieu
equation, is independent of the choice of the damping model. 

Next, we show the frequency dependence of the critical excitation
amplitude $a_{cr}$ for bare water and $d = 20 \: {\mu}{\rm m}$ in
Fig.\ \ref{fig.disp1}(b).
Both cases compare very well to the phenomenological model with
damping $\gamma_b + \gamma_s$ while we find a large discrepancy with
the prediction without surface boundary layer damping.
The bare water surface (but with our unavoidable surface
contamination) has a critical excitation amplitude that is
indistinguishable from water covered by a $d = 20 \: \mu{\rm m}$
membrane.  This proves the existence of an inextensible surface layer
on bare contaminated water, as it has the same no-slip boundary
condition at the surface as the covered water.

For the thicker membranes, the critical excitation acceleration
amplitude depends strongly on the thickness of the sheets.  In Fig.\
\ref{fig.disp5}(b) the critical excitation acceleration amplitudes
are plotted as a function of $k$ in a log-log plot, and display
remarkable scaling behavior.
The power law dependence $a_{\rm cr} \propto k^b$ has scaling
exponent $b$ comparable to 2.  The thickness dependence of $a_{cr}$
at $k = 600\:{\rm m}^{-1}$ is shown in Fig.\ \ref{fig.disp5}(c); it
is consistent with $a_{cr} \propto d^2$.

Damping factors can be understood from the dispersion relation Eq.(\
\ref{eq.disprel.1}).  By replacing $H$ with $H - \delta$ with
$\delta$ the complex frequency dependent boundary layer thickness,
$\delta = \ii \: \omega / \nu$, $\gamma_s$ follows as the imaginary
part of $\omega$ \citep{Lighthill1978}.  Similarly, endowment of the
elastic constant $E$ with an imaginary part $\propto \ii \omega E'$
(The Kelvin-Voigt model \citep{Sree2018}), leads to a damping factor
of the cover $\gamma_c$.  This factor would inherit the $k^5$
dependence on the wavenumber, and the $d^3$ dependence on the sheet
thickness.  As Fig.\ \ref{fig.disp5}(b) demonstrates, this is not the
case.  On the other hand, we have found that stretch rather than
bending dominates the dispersion relation, which leads to a weaker
dependence on $k$ and $d$ ($\propto k^3$ and $\propto d$, respectively).  

\section{Three-wave interaction}
\label{sec.three}
The dispersion relation Eq.\ \ref{eq.disprel.1}, which links
wavelength to frequency, is but one aspect of Faraday waves.  Beyond
onset, the amplitude of the waves saturates to a value which is
determined by interaction of the primary wave at $\omega = F / 2$
with daughter waves at $\omega_{1,2} = F$ which spawn nonlinearly by
the primary wave.  The wave vectors of this three-wave interaction
$\vec{k}_1 + \vec{k}_2 = \vec{k}$ must satisfy the ``energy''
condition $\omega(k_1) + \omega(k_2) = \omega(k)$. Whether this
resonance condition is possible or not depends on the shape of the
dispersion relation ($\omega$ must increase faster than linearly with
$k$).  

The mode amplitudes $B$ of Faraday waves satisfy an amplitude
equation, which we illustrate for the square pattern of Fig.\
\ref{fig.surf}(a,d), 
\be
   \frac{\d B}{\d t} = \tau^{-1} \: B - [g_0 + g(\pi/2)]\: B^3,
\label{eq.amplitude}   
\ee   
with $\tau^{-1} = \epsilon \: \gamma \: \omega$ the growth rate and
$g_0$ and the function $g(\theta)$ interaction parameters. It is a
formidable task to derive these from the Navier-Stokes equations
\citep{Vinals98,Chen99}.  Otherwise, the amplitude equation is
completely generic: there is linear growth, followed by cubic
saturation.  A square nonlinearity is absent: waves are subharmonic,
so that Eq.\ (\ref{eq.amplitude}) must be invariant under $B
\rightarrow -B$ (a time shift of one period of the driving).  In the
stationary case, $\d B/ \d t = 0$, the amplitude $B$ of a square
pattern follows as
\be
   B_\infty = \left( \frac{\epsilon \: \gamma \: \omega}
   {g_0 + g(\pi/2)} \right)^{1/2}.
\label{eq.amplitude.solution}
\ee
Incidentally, because of the localized excitation of waves, the
amplitude equation Eq.(6) in \citet{Deike2017} has a {\em quadratic}
nonlinearity.

As Eq.\ (\ref{eq.amplitude}) and Eq.\ (\ref{eq.amplitude.solution})
illustrate, the growth rate of waves should be proportional to the
scaled distance $\epsilon$ above threshold, while the amplitude of
Faraday waves saturates to a value proportional to $\epsilon^{1/2}$,
with a prefactor that depends on the details of the wave interaction.

\begin{figure}[t]
\centering
\includegraphics[scale = 0.8]{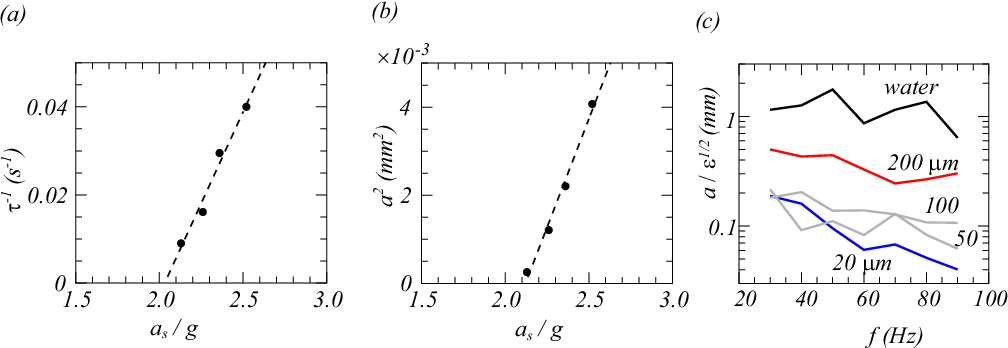}
\caption{
Illustration of the nonlinear three-wave context of Faraday waves.
For (a,b) $d = 200\: \mu{\rm m}$, and $F = 80 \: {\rm Hz}$.  Panel
(a): Amplitude growth rate $\tau^{-1}$ as a function of
(dimensionless) excitation acceleration amplitude $a_s / g$, dashed
line: fit of $\tau^{-1} \propto a_s$.  Panel (b): asymptotic (mean
squared) wave amplitude $a^2$.  Dashed line: $a^2 \propto a_s$. 
(c) Scaled asymptotic wave amplitudes $\epsilon^{-1/2} \: a$.
}
\label{fig.3w}
\end{figure}

Although the focus of the present work is not on the threshold
behavior of hydroelastic waves, some of these predictions are
illustrated in Fig.\ \ref{fig.3w}.
For a selected case, Fig.\ref{fig.3w}(a) shows the critical slowing
down at threshold $a_s = a_{cr}$ of the growth rate $\tau^{-1}
\propto \epsilon$ of viscoelastic waves, while Fig.\ref{fig.3w}(b)
shows the dependence of the asymptotic amplitude on $a_s$, which
follows $a \propto \epsilon^{1/2}$.  

Most importantly, Fig.\ref{fig.3w}(c) shows the scaled asymptotic
wave amplitudes $\epsilon^{-1/2} a$ for different thicknesses.
Comparing bare water and water covered by a sheet of $20\: \mu{\rm
m}$, we notice a decrease of almost one order of magnitude of the
amplitudes.  Since the dispersion relation and the onset of these two
cases is almost the same, this difference must be ascribed to the
change of the nonlinear three-wave interaction due to covering the
waves.

%--------------------------------------------------------------------
%--------------------------------------------------------------------

\section{Conclusion}
Although hydroelastic Faraday waves only differ from other studies
\citep{Deike2013,Deike2017,Domino2018,Biot2019} in the way of
excitation, they behave very differently.  The wavelength is
underestimated by the dispersion relation.  This discrepancy can be
quantified by introducing an additional in-plane tension in the
membrane. Although for normal in-plane tension a proportional
relation with the thickness is expected, we here find a surprising
scaling with $d$ close to $d^{3/2}$. 

We exclude a possible explanation for the significant wavelength
increase due to the wave steepness, an effect that was found by
\citet{Deike2013}, but for a clamped membrane.  We speculate that 
this nonlinear behaviour depends on the membrane boundary conditions
(clamped versus free).

Waves on bare water (but with unavoidable surface contamination)
order in squares: orthogonal one-dimensional wave crests.  Covered
waves are disordered, but with a well-defined wavelength.  We
quantify their anisotropy in several ways, with the conclusion that a
thick cover supports slightly stronger anisotropy.  

The damping of hydroelastic Faraday waves depends strongly on the
cover thickness.  The damping of waves on bare, unavoidably
contaminated, water can be predicted by the bulk damping and the
boundary layer dissipation when an inextensible surface layer (i.e. a
no-slip condition) is assumed. The results for wave damping under the
thinnest membrane, $d = 20 \: \mu{\rm m}$, are identical to those of
bare water.  For thicker membranes the threshold amplitude grows
proportional to $k^2$ and proportional to $d^2$ with increasing
wavenumber and thickness, respectively.
Lastly, the presence of the membrane has a large effect on the wave
amplitude, an effect we ascribe to nonlinear wave interactions.

\begin{acknowledgments}
We thank Jerry Westerweel who proposed the size reduction of floats
on ocean waves to the laboratory scale.
We also thank Sebastian Schreier and Peter Wellens for discussions
and help with the laboratory facilities.  
This study was funded by the Dutch Research Council (NWO) under the
Grant Number 19002 (``FlexFloat'') in the Open Technology Program
(OTP). The authors are grateful for the technical support by
S.Tokg\"{o}z and F.J. Sterk.
\end{acknowledgments}

%.......................................................... appendix
%....................................................................
\appendix* % the star omits numbering
\label{sec.appendix}
\section{Dispersion and critical excitation amplitude using silicon oil}
\begin{figure}[h]
\centering
\includegraphics[scale = 0.8]{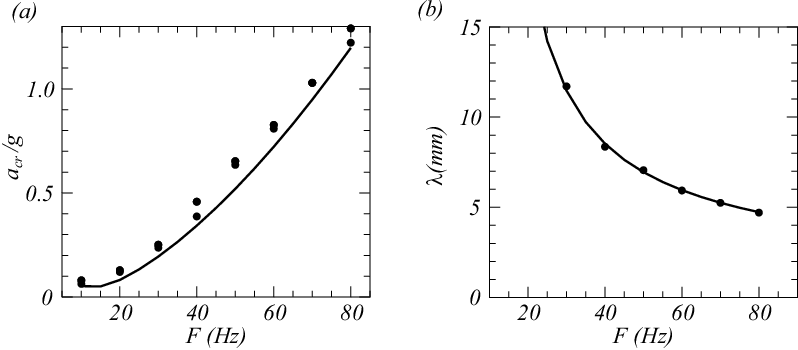}
\caption{(a) Dots: critical dimensionless excitation
amplitude $a_{cr}$. Notice that measured values are overestimated due
to the critical slowing down at hreshold.  (b): Dispersion relation.
Dots are measurements, lines are the prediction of \citet{Kumar94}. }
\label{fig.silicon}
\end{figure}

We verify our experimental setup, the measurement of the dispersion
relation and the measurement of the critical excitation amplitude
using (uncovered) silicon oil. Contrary to water, silicon oil ($\rho
= 910\: {\rm kg}{\rm m}^{-3}$, $\eta = 4.55 \: {\rm m Pa s}$, $\sigma =
1.97 \times 10^{-2}\: {\rm N}{\rm m}^{-1}$) has a reproducible
surface tension which is not sensitive to contamination.  As Fig.\
\ref{fig.silicon} shows, we find excellent agreement with the
theoretical predictions of \citet{Kumar94}) for the onset amplitude
and the dispersion relation.

%--------------------------------------------------------------------
%--------------------------------------------------------------------

%....................................................................
%\bibliography{faraday}

\newpage
\end{document}